\documentclass[copyright,creativecommons]{eptcs}

\usepackage{breakurl}             % Not needed if you use pdflatex only.
\usepackage{amssymb}
\usepackage{amsmath}
\usepackage{./proof}

\title{A coinductive semantics of the Unlimited Register Machine}
\author{Alberto Ciaffaglione
\institute{Dipartimento di Matematica e Informatica \\ Universit\`a di Udine, Italia}
\email{alberto.ciaffaglione@uniud.it}
}
\def\titlerunning{A coinductive semantics of the Unlimited Register Machine}
\def\authorrunning{A. Ciaffaglione}

\newcommand{\proof}{{\scshape\noindent Proof.\quad}}

\newcommand {\eg}{\textit{e.g}.}
\newcommand {\ie}{\textit{i.e}.}
\newcommand {\wrt}{\textit{w.r.t.}}

\newcommand{\coq}{\textsf{Coq}}

\newcommand{\cic}{\emph{Calculus of (Co)Inductive Constructions}}
\newcommand{\cccoind}{\ensuremath{\text{CC}^\text{(Co)Ind}}}

\newcommand{\N}{I\!\!N}

\newcommand{\rangeiM} {{\iota \in [1..m]}}
\newcommand{\rangeiMinf} {{\iota \in [1..\infty]}}
\newcommand{\rangeiMtwoinf} {{\iota \in [2..\infty]}}
\newcommand{\rangeiMtwo} {{\iota \in [2..m]}}
\newcommand{\rangeiN} {{\iota \in [1..n]}}
\newcommand{\rangeiRho} {{\iota \in [1..\rho(U)]}}

\newtheorem{thm}{Theorem}[section]
\newtheorem{adef}[thm]{Definition}

\newtheorem{cjt}[thm]{Conjecture}

\begin{document}
\maketitle

\begin{abstract}
  We exploit (co)inductive specifications and proofs to approach the
  evaluation of low-level programs for the \emph{Unlimited Register
    Machine (URM)} within the \coq\ system, a proof assistant based on
  the \cic\ type theory.
  Our formalization allows us to certify the implementation of partial
  functions, thus it can be regarded as a first step towards the
  development of a workbench for the formal analysis and verification
  of both converging and diverging computations.
\end{abstract}

%%%%%%%%%%%%%%%%%%%%%%%%%%%%%%%%%%%%%%%%%%%%%%%%%%%%%%%%%%%%%%%%

\section{Introduction}
\label{sec:intro}

In this paper we report and discuss a formalization of the
\emph{Unlimited Register Machine} (URM) and its semantics within the
\cic\ (\cccoind).

The URM is a mathematical idealisation of a computer, one of the
formal approaches to characterize the intuitive ideas of computability
and decidability \cite{cutland}.
Programs for the URM are low-level, essentially assembly-like, and
their execution gives rise to both converging and diverging
computations.
This is a typical situation where it is required to define and reason
about \emph{circular, potentially infinite} objects and concepts, \ie\
systems with infinitely many states.  Since structural induction
trivially fails on these systems, one may resort to stronger
approaches, such as, among other ones, \emph{coinduction}.

Coinductive principles can be stated and exploited in different
settings.
From a \emph{set-theoretical} standpoint coinduction arises when
objects are viewed as \emph{maximal fixed-points} of monotone
operators, whereas the \emph{categorical} approach is developed
through \emph{(final) coalgebras}.
To develop the present work, we settle within the \emph{logical}
system of \emph{Intuitionistic Type Theory}.

Actually, in intuitionistic type theory infinite objects are managed
through \emph{coinductive types}: these, roughly speaking, are
collections of elements whose construction requires an infinite
numbers of steps.
In particular, a handy technique for dealing with coinductive
definitions and proofs within \cccoind\ was introduced by Coquand
\cite{coquand93} and refined by Gim\'enez \cite{gimenez94}.  Although
providing a limited form of coinduction, such an approach is
particularly appealing, because \emph{proofs} carried out by
coinduction are accommodated as any other infinite, coinductively
defined object.
Remarkably, such a technique is mechanised in the system \coq\
\cite{coq}: this, one among the rare interactive environments that
implement coinductive definition and proof principles, is an
appreciated proof assistant, due to the fact that the automatization
and the interaction with the user are well-balanced.

In this paper we formalize the URM and its semantics from the point of
view of the \emph{program certification}.
In our opinion, such an encoding within a coinductive formal system,
such as \cccoind, has several benefits.  First it is interesting
\emph{per se}, as experiments about the encoding of computability
models are still lacking.
Then it may be valuable in education, by giving the opportunity to
undergraduate students (computability is actually a basic computer
science course) to experiment with non-standard (\ie\ coinductive)
tools within a concrete, relatively simple application.
Further it might be useful in the area of program transformations,
because the formal treatment of low-level languages is mandatory to
certify components of programming languages, such as type-checkers,
interpreters, and compilers.
Last but not the least, the present, novel theoretical case study
witnesses the broad applicability of coinduction as a verification
technique on infinite-state systems and the significance of its
mechanisation.

Besides the points mentioned above, we claim that the originality of
this paper relies also on the presentation of the encoding, which is
illustrated and discussed without showing \coq\ code, but via the more
abstract level of \cccoind\ (in any case, the \coq\ code is available
to the interested reader at the web page of the author
\cite{alberto:url}), thus providing the reader with an extra pedagogic
value.

In the next section we illustrate coinduction within \cccoind; then in
the following four sections we develop the formalization of the URM,
dealing with programs, computations and functions; finally we discuss
directions for further investigations in the light of what we achieve
and of related work.

%%%%%%%%%%%%%%%%%%%%%%%%%%%%%%%%%%%%%%%%%%%%%%%%%%%%%%%%%%%%%%%%

\section{Coinduction in \cccoind}
\label{sec:coind}

The formal treatment of infinite objects and concepts is supported by
\cccoind\ via the mechanism of coinductive types. These, by providing
the user with a limited form of recursion, allow the formalization and
the management of infinite data and infinite proofs.

First of all, one may define concrete, \emph{infinite} objects (\ie\
data) as elements of \emph{coinductive types}, which are fully
described by a set of \emph{constructors}\footnote{The constructors
  must respect a \emph{strict positivity constraint} condition to
  guarantee the reduction termination of the calculus.}.  From a pure
logical point of view, the constructors can be seen as
\emph{introduction rules}; these are interpreted coinductively, \ie\
they are applied infinitely many times, hence the type being defined
is inhabited by infinite objects:
\[
\begin{array}{c}
  \infer[(0S)_{\infty}]
  {0{:}s \in S}
  {s \in S}
  \qquad
  \infer[(1S)_{\infty}]
  {1{:}s \in S}
  {s \in S}
\end{array}
\]
In this case we have formalized infinite sequences, \ie\
\emph{streams}, of bits, a coinductive type we name $S$.
Optionally, coinductive types may contain finite objects too, that is,
\emph{potentially} infinite objects; in such a case also
\emph{constant} constructors, besides the recursive ones, have to be
declared:
\[
\begin{array}{llll}
  \infer[(0L)]
  {0 \in L}
  {}
  & \quad
  \infer[(1L)]
  {1 \in L}
  {}
  & \quad
  \infer[(0L)_{\infty}]
  {0{:}l \in L}
  {l \in L}
  & \quad
  \infer[(1L)_{\infty}]
  {1{:}l \in L}
  {l \in L}
\end{array}
\]
So doing, we have defined $L$, the type of sequences of both finite
and infinite length, \ie\ \emph{lazy lists}, of bits.

Once a new coinductive type is defined, the system provides
automatically the \emph{destructors}, \ie\ an extension of the native
pattern-matching capability, to \emph{consume} the elements of the
type itself.  Therefore, coinductive types can also be viewed as the
\emph{largest} collection of objects closed \wrt\ the destructors.

Consistently with this intuition, the destructors \emph{cannot} be
used for defining functions by recursion on coinductive types, because
their elements cannot be consumed down to a constant case.  The
natural way to allow self-reference is to consider the dual
perspective of \emph{building} individual, constant elements in
coinductive types. Such a goal can be fullfilled through \emph{lazy
  corecursive} functions:
\[
\begin{array}{lcl}
zeros & \triangleq  & 0 {:} zeros \\
odd(s) & \triangleq  & \textrm{match $s$ with $a{:}b{:}s'$ $\Rightarrow$ $a{:}odd(s')$} \\
even(s) & \triangleq  & \textrm{match $s$ with $a{:}b{:}s'$ $\Rightarrow$ $b{:}even(s')$} \\
merge(s,t) & \triangleq  & \textrm{match $s$ with $a{:}s'$ $\Rightarrow$
                                   match $t$ with $b{:}t'$ $\Rightarrow$ $a{:}b{:}merge(s',t')$}
\end{array}
\]
Corecursive functions produce infinite objects and may have any type
as domain (note that in the last three definitions we have applied the
\emph{match} destruction operation on a parameter of the domain).
Infinite objects are not unfolded, unless their components are
explicitly needed, ``on demand'', by a destruction operation.
Therefore, to prevent the evaluation of corecursive functions from
infinitely looping, their definition must satisfy a \emph{guardedness
  condition}: every corecursive call has to be guarded by at least one
constructor, and by nothing but constructors\footnote{Syntactically,
  the constructors guard the recursive call ``on the left''.}.
This way of regulating the implementation of corecursion captures the
intuition that infinite objects are built via the iteration of an
initial step.

Given a concrete coinductive type (such as $S$ and $L$ above), no
proof principle can be automatically generated by the system: in fact,
proving properties about infinite objects requires the potential of
building \emph{proofs} which are infinite as well!
What is needed is the design of \emph{ad-hoc} coinductive
\emph{predicates}, \ie\ coinductive \emph{propositions}, which are
actually inhabited by such \emph{infinite proofs}\footnote{This
  distinction between concrete objects and proofs points out that sets
  inhabited by concrete objects have \emph{computational} content,
  whereas predicates inhabited by proofs carry \emph{logical}
  information.}.  The traditional example is point-wise equality (also
known as \emph{bisimilarity}), that we define on streams and name
$\simeq \; \subseteq S \times S$:
\[
\begin{array}{c}
  \infer[(\simeq)_{\infty}]
  {b{:}s \simeq b{:}t}
  {b {\in} \{0,1\} \quad s \simeq t}
\end{array}
\]
Two streams are bisimilar if we can \emph{observe} that they have
equal heads and recursively, \ie\ \emph{coinductively}, their tails
are bisimilar. Once this new predicate is defined, the system provides
the corresponding \emph{proof principle}, to carry out proofs about
bisimilarity: such a tool, named \emph{guarded induction} principle
\cite{coquand93,gimenez94}, is particularly appealing in a context
where proofs are managed as any other infinite object.

In fact, a proof by guarded induction is just an infinite object built
by lazy corecursion (hence it must respect the same guardedness
constraint that lazy corecursive functions have to).
Remarkably, the mechanization of the guarded induction principle
provides a handy technique for the construction of infinite proofs,
which can be carried out interactively through the \texttt{cofix}
tactic\footnote{A tactic is a command to solve a goal or decompose it
  into simpler goals.}. This tactic allows to build infinite proofs as
\emph{infinitely regressive} proofs, by assuming the thesis as an
extra hypothesis and using it carefully later, provided its
application is guarded by constructors.
This ``internal'' approach is very direct, compared to the traditional
techniques based on bisimulations, because the proofs do not need to
be exhibited beforehand, but can be built incrementally via tactics.

To illustrate the support provided by the \texttt{cofix} tactic, we
pick out the following coinductive property:
\[
\forall s {\in} S.\ merge(odd(s),\ even(s)) \simeq s
\]
We prove this proposition by mimicking the top-down proof practice of
\cccoind. First, the coinductive hypothesis is assumed among the
hypotheses and the stream $s$ is destructed two times into
$a{:}b{:}t$; then the corecursive functions $odd$, $even$ and $merge$,
in turn, may perform a computation step; finally the constructor
$(\simeq)_{\infty}$ is applied twice.
In the end, we have reduced the goal to prove $merge(odd(t),\ even(t))
\simeq t$, a proposition which is an instance of the coinductive
hypothesis.
Therefore one is eventually allowed to exploit the coinductive
hypothesis itself, whose application is now guarded by the constructor
$(\simeq)_{\infty}$. The application of the coinductive hypothesis
completes the proof, and intuitively has the effect of repeating ad
infinitum the explicit, initial proof segment, thus realizing the
``and so on forever'' motto.

To avoid ambiguity with genuine induction, we say that the proof has
been performed by \emph{structural coinduction} on the derivation.
The whole proof may be displayed in natural deduction
style\footnote{As usual, local hypotheses are indexed with the rules
  they are discharged by.} as follows:
\[
\infer[(1)]
  {\forall s {\in} S.\ merge(odd(s),\ even(s)) \simeq s}
    {\infer[(introduction)]
       {\forall s {\in} S.\ merge(odd(s),\ even(s)) \simeq s}
       {\infer[(destruction)]
         {merge(odd(s),\ even(s)) \simeq s}
         {\infer[(computation{:}\ odd,\ even)]
           {merge(odd(a{:}b{:}t),\ even(a{:}b{:}t)) \simeq a{:}b{:}t}
           {\infer[(computation{:}\ merge)]
             {merge(a{:}odd(t),\ b{:}even(t)) \simeq a{:}b{:}t}
             {\infer[(\simeq)_{\infty}]
               {a{:}b{:}merge(odd(t),\ even(t)) \simeq a{:}b{:}t}
               {[merge(odd(t),\ even(t)) \simeq t]_{(1)}}}}}}}
\]

To conclude, we observe that, as the reader may imagine, there exist
several \emph{semantically} productive\footnote{Productivity is the
  power of a function call to produce data, which is undecidable.},
but \emph{syntactically} non-guarded functions (and proofs) that
cannot be accepted by \cccoind, because the automated check is not
sophisticated enough.
Particular effort is put in fact by the community into the goal of
extending the expressive power of guarded corecursion
\cite{gimenez98,digianantonio-miculan02,bertot:reccorec}.  At the
moment, we can say that \cccoind\ has made a lot of progress, but
there are still problematic issues on the carpet.

%%%%%%%%%%%%%%%%%%%%%%%%%%%%%%%%%%%%%%%%%%%%%%%%%%%%%%%%%%%%%%%%

\section{The Unlimited Register Machine}
\label{sec:urm}

The Unlimited Register Machine (URM) is a mathematical idealisation of
a computer, one among the frameworks proposed to set up a formal
characterisation of the intuitive ideas of effective computability and
decidability. It is equivalent to the alternative approaches, \eg\
Turing machines, and particulary valued for its simplicity.  We work
here with the URM formulation introduced by Cutland \cite{cutland}, a
slight variation of a machine first conceived by Shepherdson and
Sturgis \cite{urm63}.

\paragraph{\textbf{Registers and instructions.}}

The URM has an \emph{infinite} number of \emph{registers} $R_1, R_2,
\ldots$ containing natural numbers $r_1, r_2, \ldots$ which may be
altered by \emph{instructions}. These are of four kinds and have the
following intended meaning ($r \to R$ represents the loading of the
natural value $r$ in the register $R$):
\[
\begin{array}[t]{lclcl}
Z(i)      & \triangleq & \textrm{Zero}      & : & 0 \to R_i\\
S(i)      & \triangleq & \textrm{Successor} & : & r_i + 1 \to R_i\\
T(i,j)    & \triangleq & \textrm{Transfer}  & : & r_i \to R_j\\
J(i,j,k)  & \triangleq & \textrm{Jump}      & : & \textrm{if $r_i {=} r_j$ then proceed from the $k$th instruction} \\
          &            &                    &   & \phantom{\textrm{if $r_i {=} r_j$ }} 
                                     \textrm{else proceed from the next instruction}\\
\end{array}
\]

\paragraph{\textbf{Programs and computations.}}

A \emph{program} for the URM is a finite, non-empty sequence of
instructions.

When provided with a program $P$ and a(n \emph{initial) configuration}
(\ie\ a \emph{finite}, non-empty sequence of natural numbers $r_1,
r_2, \ldots, r_m$ in the registers $R_1, R_2, \ldots,
R_m$)\footnote{Despite the number of the registers being infinite, any
  program $P$ is finite, so there exists a maximal register index $m
  {=} \rho(P)$, depending on $P$, such that $R_m$ is affected by the
  instructions in $P$.  Hence $r_1, r_2, \ldots, r_m$ is equivalent to
  $r_1, r_2, \ldots, r_m, 0, 0, \ldots$}, the URM performs a
\emph{computation}: this means starting from the first instruction in
$P$ and obeying the instructions sequencially (unless a Jump is
encountered), thus altering at any step the content of the registers
as prescribed by the instructions.

The computation \emph{stops}, or \emph{converges}, if and only if
there is no next instruction; when this is the case, the number $r$
stored in $R_1$ in the \emph{final} configuration is regarded as the
output of the computation, and this is written $P(r_1, r_2, \ldots,
r_m) \!\downarrow\!  r$.
On the other hand, due to the looping back via the Jump instruction,
there are computations that \emph{never stop}, or \emph{diverge},
which is written $P(r_1, r_2, \ldots, r_m) \!\uparrow$.

\paragraph{\textbf{Formalization in \cccoind.}}

The encoding of the basic URM structures in \cccoind\ is
straightforward, because both configurations and programs are simply
finite, non-empty sequences of components, which we formalize by means
of inductive datatypes (the $\N$ represents the natural numbers):
$$
\begin{array}[h]{lclcll}
  Loc & : & i,j    & \in & \N^+ {=} \N {-} \{0\} & \quad \textrm{register index} \\
  Val & : & r      & \in & \N & \quad \textrm{register content} \\
  Cgn & : & \sigma & ::= & (\iota {\mapsto} r_\iota)^\rangeiM & \quad \textrm{list-configuration} \\
  \\
  PC  & : & k,h & \in & \N & \quad \textrm{program counter} \\
  Inst& : & I   & \in & \{Z(i),\ S(i),\ T(i,j),\ J(i,j,k)\} & \quad \textrm{instruction} \\
  Pgm & : & U,V & ::= & \langle \iota {\mapsto} I_\iota\rangle^\rangeiN & \quad \textrm{program} \\
\end{array}
$$
An alternative encoding of configurations can be given via
\emph{infinite} sequences, \ie\ coinductive datatypes:
$$
\begin{array}[h]{lclcll}
  Cgn_{\infty} & : & \sigma_{\infty} & ::= & (\iota {\mapsto} r_\iota)^\rangeiMinf & \quad \textrm{stream-configuration} \\
\end{array}
$$

\paragraph{\textbf{Adequacy (I).}}

We start to address now the faithfulness of our encoding of the URM,
by comparing Cutland's formulation and our formalization in \cccoind.
First, we observe that the syntax of our instructions (and therefore
of programs) coincide with Cutland's one.
Then, two technical points have to be considered: about the
convergence of computations, and about the encoding of configurations.

The ``natural'' way for the program $U {=} {I_1, I_2, \ldots, I_n}$ to
stop is that the program counter is set eventually to $n{+}1$; though,
a Jump instruction could set it to an index greater than $n{+}1$.
Cutland actually confines his attention to the programs that
invariably stop because the next instruction should be $I_{n+1}$.  We
adopt a similar convention here, with the difference that we use the
index $0$ in place of $n{+}1$: these kinds of programs, the sole we
will be considering from now on, are said to be \emph{in standard
  form}.

\begin{adef}\;(Standard form)\\
  A program $U {=} \langle \iota {\mapsto} I_\iota\rangle^\rangeiN$ is
  \emph{in standard form} if, for every $J(i,j,k) {\in} U$, $k {\leq}
  n$ holds. 
\end{adef}

As far as the formalization of configurations is concerned, it is
apparent that our stream-configurations (\ie\ the datatype
$Cgn_{\infty}$) correspond to infinite sequences of registers in the
original URM.

By working \emph{on paper}, on the one hand, Cutland is naturally
allowed to define configurations as finite, starting segments of such
infinite sequences of registers: in fact, by inspecting a given
program $P$, one can pick out $\rho(P)$, the maximal register index
affected by the instructions in $P$.
In this way the working space available to the computation under $P$
may be restricted to the configuration $r_1, r_2, \ldots,
r_{\rho(P)}$.

On the other hand, working \emph{formally} within \cccoind\ requires extra care.
First we observe that our list-configurations (\ie\ the datatype
$Cgn$) correspond to the above Cutland configurations $r_1, r_2,
\ldots, r_{\rho(P)}$.
Nevertheless, list-configurations bring a drawback: if one wants to
reason formally on them, it is required to consider only programs that
respect the working space they make available\footnote{In a sense,
  this means to provide in advance with the maximal register index
  $\rho(U)$, given a program $U$.}.
That is, programs and list-configurations can be soundly coupled just
if the programs contain ``good'' pointers (\ie\ indexes) to the
configurations themselves, a constraint that can be viewed as a kind
of \emph{compatibility} concept.

\begin{adef}\;(Compatibility) A program $U$ and a list-configuration
  $\sigma {=} (\iota {\mapsto} r_\iota)^{i \in [1..m]}$ are
  \emph{compatible} ($\sigma \models U$) if $U$ is in standard form
  and, for every $Z(i)$, $S(i)$, $T(i,j)$, $J(i,j,k) {\in} U$, $i,j
  {\in} [1..m]$ holds. 
\end{adef}

%%%%%%%%%%%%%%%%%%%%%%%%%%%%%%%%%%%%%%%%%%%%%%%%%%%%%%%%%%%%%%%%

\section{Abstract computation}
\label{sec:abstract}

In this section we bootstrap the semantics of the URM, by extending in
a modular way the formalization introduced so far; note that, from now
on, we will use the terminology ``configuration'' to refer to the
encoding in \cccoind\ itself (\ie\ either the finite-list datatype
$Cgn$ or the infinite-stream datatype $Cgn_{\infty}$).

It is apparent that the concept of \emph{convergence} of computations
can be relativised \wrt\ configurations: there are actually programs
that always stop and programs that never stop (whatever configuration
is coupled to them) and programs that either converge or diverge
depending on the initial configuration.
Clearly, the divergence is caused by the presence of \emph{infinite
  loops} in the progress of computation: to deal formally with the
execution of programs we have then to manage an infinite-state system,
a scenario which may benefit from the use of the \emph{coinduction} as
a specification and proof principle.

In this section we focus just on a restricted, basic notion of
computation: in fact, from the point of view of the termination, the
only essential instruction is the $Jump$ instruction, which has the
capability to \emph{separate} converging computations from diverging
ones.  Hence we consider here programs that contain only Jump
instructions, \ie\ \emph{abstract} programs; this preliminary
investigation allows us to focus on the object system from a cleaner
perspective, to be exploited in the following.

Noticeably, it is not possible to cope with the semantics of URM
programs by using a unique, \emph{potentially} coinductive computation
concept (see Section \ref{sec:coind}): a faithful encoding has
actually to reflect the separation between converging and diverging
computations, through two different judgments.
Therefore, using in this case \emph{finite} (\ie\ list)
configurations, the semantics of abstract URM programs can be
described by the \emph{inductive} $cp_{j+}$ and the \emph{coinductive}
$cp_{j{\infty}}$ predicates, whose arity is $Pgm \times Cgn \times
PC$.

\begin{adef}\label{abs}\;(Abstract evaluation) 
  Let $A {=} \langle \iota {\mapsto} I_\iota\rangle^\rangeiN$ and
  $\sigma {=} (\iota {\mapsto} r_\iota)^\rangeiM$ be an abstract
  program and a configuration such that $\sigma \models A$, and let $h
  \in [1..n]$ and $I_h {=} J(i,j,k)$.  Then, $cp_{j+}$ is defined by
  the first four rules, interpreted inductively, and $cp_{j\infty}$ by
  the last two rules, interpreted coinductively:
\[
\begin{array}{cc}
  \infer[(f{\cdot}l)_+]
  {cp_{j+}(A, \sigma, h)}
  {h {=} n \quad r_i {\neq} r_j}
  & \qquad
  \infer[(t{\cdot}l)_+]
  {cp_{j+}(A, \sigma, h)}
  {k {=} 0 \quad r_i {=} r_j}
  \\
  \\
  \infer[(f{\cdot}r)_+]
  {cp_{j+}(A, \sigma, h)}
  {cp_{j+}(A, \sigma, h{+}1) \quad h {<} n \quad  r_i {\neq} r_j}
  & \qquad
  \infer[(t{\cdot}r)_+]
  {cp_{j+}(A, \sigma, h)}
  {cp_{j+}(A, \sigma, k) \quad k {\neq} 0 \quad r_i {=} r_j}
  \\
  \\
  \infer[(f{\cdot}r)_{\infty}]
  {cp_{j\infty}(A, \sigma, h)}
  {cp_{j\infty}(A, \sigma, h{+}1) \quad h {<} n \quad  r_i {\neq} r_j}
& \qquad
  \infer[(t{\cdot}r)_{\infty}]
  {cp_{j\infty}(A, \sigma, h)}
  {cp_{j\infty}(A, \sigma, k) \quad k {\neq} 0 \quad r_i {=} r_j}
\end{array}
\]
\end{adef}

At the moment, our goal is to capture just the progress of the
\emph{control flow}, with the computation that may proceed from a
generic instruction of a program.
Specifically, the intended meaning of the judgments
$cp_{j+}(A,\sigma,h)$ and $cp_{j{\infty}}(A,\sigma,h)$ is that the
computation under the abstract program $A$ with the configuration
$\sigma$ and by starting from the $h$th instruction of $A$,
\emph{converges} and \emph{diverges}, respectively.

More in detail, the coinductive predicate asserts that the computation
loops: that is, by starting from the instruction $I_h$, there exists
an instruction $I_q$ which can be reached from $I_h$ and such that,
afterwards, the control flow comes again at $I_q$ after a non-zero,
finite number of steps.  Hence, the divergence is grasped via the
predicate $cp_{\infty}$ by the coinduction proof principle motto
(``and so on forever'').

We remark that, since URM programs are not structured, we have to
embed in the encoding some other ``structuration'' criterium; in fact,
the design of the predicates has been directly inspired by the
\emph{number of evaluation steps} implicit amount. Thus we have
defined two atomic rules for $cp_{j+}$ (the evaluation stops in one
step), when either the current one is the last instruction and the
Jump condition is false, or the current Jump condition is true and the
instruction tells to jump out of the program. The extra rules are
recursive, and address how an evaluation step is carried out within a
converging computation (predicate $cp_{j+}$) and a diverging one
(predicate $cp_{j{\infty}}$), again inspecting by cases the Jump
condition.

Another important choice to be pointed out is that we have modeled the
evaluation from a particular perspective, \ie\ for using the
judgments, according to \coq's top-down proof practice, to
\emph{execute} specific programs. This ``algorithmic'' approach is
motivated by the fact that we are interested in experimenting the
certification of concrete programs; this is a preliminary step that
pinpoints further investigations, such as the development of the
metatheory of the URM or the advanced issues addressed by Leroy and
Grall \cite{leroy-grall09}.  We are conscious that these more
ambitious tasks could require the introduction of new versions of the
evaluation concept, to be related to the ones we have formalized up to
date.

We notice, finally, that a fragment of the encoding of the evaluation
judgments, which is common to all the rules, has not been displayed in
the rules themselves, but has been collected within the hypotheses of
the Definition \ref{abs}: such a part of the formalization has to cope
with the compatibility between programs and finite configurations, an
overhead that we have discussed in the previous section.

In the end, using our machinery we can manage termination and
divergence of computations under abstract URM programs parameterically
\wrt\ non-mutable configurations, as follows.

\begin{adef}\label{abstracteval}\;(Converging and diverging abstract evaluation) 
  Let $A$ and $\sigma$ be an abstract program and a configuration such
  that $\sigma \models A$. The computation under $A$ with $\sigma$
  \emph{converges} and \emph{diverges} when:
\[
\begin{array}{lcl}
  stop_j(A,\sigma) & \triangleq & cp_{j+}(A, \sigma, 1)
  \\
  loop_j(A,\sigma) & \triangleq & cp_{j\infty}(A, \sigma, 1)
\end{array}
\]
\end{adef}

As an example, let us consider the abstract program $B {\triangleq}
\langle 1 {\mapsto} J(1,2,2),\ 2 {\mapsto} J(1,2,2) \rangle$. We can
prove that the computation under $B$ with the configuration $\sigma
{\triangleq} (1 {\mapsto} 0,\ 2 {\mapsto} 1)$ converges, while it
diverges with $\tau {\triangleq} (1 {\mapsto} 0,\ 2 {\mapsto} 0)$;
both the proofs are immediate, the second one is by
coinduction\footnote{As discussed in Section \ref{sec:coind}, the
  proofs are displayed in natural deduction style and have to be read
  from the bottom.}:
\[
\begin{array}{c}
  \infer[(f{\cdot}r)_+]
  {cp_{j+}(B, \sigma, 1)}
  {r_1 {=} 0 {\neq} 1 {=} r_2 & \infer[(f{\cdot}l)_+]{cp_{j+}(B, \sigma, 2)}
                                                     {r_1 {=} 0 {\neq} 1 {=} r_2}}
  \qquad
  \infer[(t{\cdot}r)_{\infty}]
  {cp_{j\infty}(B, \tau, 1)}
  {2 {\neq} 0 & r_1 {=} 0 {=} r_2 & \infer[(t{\cdot}r)_{\infty} (1)]{cp_{j\infty}(B, \tau, 2)}
                                                                    {2 {\neq} 0 & r_1 {=} 0 {=} r_2 & [cp_{j\infty}(B, \tau, 2)]_{(1)}}}
\end{array}
\]

A more sensible approach would allow to manage \emph{variable}
configurations, such as $\mu {\triangleq} (1 {\mapsto} m,\ 2 {\mapsto}
n)$.
In that case, the Definition \ref{abstracteval} should be more
involved, by including a premise to \emph{constrain} the content of
the configuration at hand. So doing, one could prove more general
assertions, such as \eg\ $(m {\neq} n) \Rightarrow cp_{j+}(B, \mu, 1)$
and $(m {=} n) \Rightarrow cp_{j\infty}(B, \mu, 1)$.
Though, we prefer to postpone such versions of convergence and
divergence to the next section, where we will address the full URM
instruction suite.

%%%%%%%%%%%%%%%%%%%%%%%%%%%%%%%%%%%%%%%%%%%%%%%%%%%%%%%%%%%%%%%%

\section{Full computation}
\label{sec:full}

We extend now our formalism to deal with the full URM, by adopting
\emph{infinite} (\ie\ stream) configurations, because these allow to
dispose of the compatibility between programs and configurations
themselves (as argued in Sections \ref{sec:urm} and
\ref{sec:abstract}). Note that the results we get are independent from
the particular encoding of configurations (in fact, at the end of this
section we will relate formally finite and infinite configurations to
each other, by addressing the adequacy of the whole formalization).

Actually, the computation under URM programs is captured by the more
involved inductive predicate $cp_+$, with arity $Pgm \times
Cgn_{\infty} \times PC \times Cgn_{\infty}$, and the coinductive
predicate $cp_{\infty}$, with arity $Pgm \times Cgn_{\infty} \times
PC$, which describe \emph{both} the control flow \emph{and} its effect
on configurations.

\begin{adef}\label{eval}\;(Evaluation) 
  Let $U {=} \langle \iota{\mapsto} I_{\iota}\rangle^\rangeiN$ and
  $\sigma_{\infty} {=} (\iota{\mapsto} r_{\iota})^\rangeiMinf$ be a
  program and a configuration, and let $h \in [1..n]$. We assume that
  $I_h {=} J(i,j,k)$ in the Jump rules (those labelled $(j{-})$), $I_h
  {=} Z(i)$ in the Zero rules, $I_h {=} S(i)$ in the Successor rules,
  and $I_h {=} T(i,j)$ in the Transfer rules.

  Then, $cp_+$ is defined by the following rules, interpreted
  inductively:
\[
\begin{array}{cc}
  \infer[(jf{\cdot}l)_+]
  {cp_+(U, \sigma_{\infty}, h, \sigma_{\infty})}
  {h {=} n \quad r_i {\neq} r_j}
&  \
  \infer[(jf{\cdot}r)_+]
  {cp_+(U, \sigma_{\infty}, h, \tau_{\infty})}
  {cp_+(U, \sigma_{\infty}, h{+}1, \tau_{\infty}) \quad h {<} n \quad r_i {\neq} r_j}
\\
\\
  \infer[(jt{\cdot}l)_+]
  {cp_+(U, \sigma_{\infty}, h, \sigma_{\infty})}
  {k {=} 0 \quad r_i {=} r_j}
& \
  \infer[(jt{\cdot}r)_+]
  {cp_+(U, \sigma_{\infty}, h, \tau_{\infty})}
  {cp_+(U, \sigma_{\infty}, k, \tau_{\infty}) \quad k {\neq} 0 \quad r_i {=} r_j}
\\
\\
  \infer[(z{\cdot}l)_+]
  {cp_+(U, \sigma_{\infty}, h, \tau_{\infty})}
  {h {=} n \quad \tau_{\infty} {=} zr(\sigma_{\infty},i)}
& \
  \infer[(z{\cdot}r)_+]
  {cp_+(U, \sigma_{\infty}, h, \tau_{\infty})}
  {cp_+(U, \sigma_{\infty}', h{+}1, \tau_{\infty}) \quad h {<} n \quad \sigma_{\infty}' {=} zr(\sigma_{\infty},i)}
\\
\\
  \infer[(s{\cdot}l)_+]
  {cp_+(U, \sigma_{\infty}, h, \tau_{\infty})}
  {h {=} n \quad \tau_{\infty} {=} sc(\sigma_{\infty},i)}
& \
  \infer[(s{\cdot}r)_+]
  {cp_+(U, \sigma_{\infty}, h, \tau_{\infty})}
  {cp_+(U, \sigma_{\infty}', h{+}1, \tau_{\infty}) \quad h {<} n \quad \sigma_{\infty}' {=} sc(\sigma_{\infty},i)}
\\
\\
  \infer[(t{\cdot}l)_+]
  {cp_+(U, \sigma_{\infty}, h, \tau_{\infty})}
  {h {=} n \quad \tau_{\infty} {=} mv(\sigma_{\infty},i,j)}
& \
  \infer[(t{\cdot}r)_+]
  {cp_+(U, \sigma_{\infty}, h, \tau_{\infty})}
  {cp_+(U, \sigma_{\infty}', h{+}1, \tau_{\infty}) \quad h {<} n \quad \sigma_{\infty}' {=} mv(\sigma_{\infty},i,j)}
\end{array}
\]
And $cp_{\infty}$ is defined by the following rules (a superset of
those for $cp_{j\infty}$), interpreted coinductively:
\[
\begin{array}{c}
  \infer[(jf{\cdot}r)_{\infty}]
  {cp_{\infty}(U, \sigma_{\infty}, h)}
  {cp_{\infty}(U, \sigma_{\infty}, h{+}1) \quad h {<} n \quad r_i {\neq} r_j}
\qquad
  \infer[(jt{\cdot}r)_{\infty}]
  {cp_{\infty}(U, \sigma_{\infty}, h)}
  {cp_{\infty}(U, \sigma_{\infty}, k) \quad k {\neq} 0 \quad r_i {=} r_j}
\\
\\
  \infer[(z{\cdot}r)_{\infty}]
  {cp_{\infty}(U, \sigma_{\infty}, h)}
  {cp_{\infty}(U, \tau_{\infty}, h{+}1) \quad h {<} n \quad \tau_{\infty} {=} zr(\sigma_{\infty},i)}
\qquad
  \infer[(s{\cdot}r)_{\infty}]
  {cp_{\infty}(U, \sigma_{\infty}, h)}
  {cp_{\infty}(U, \tau_{\infty}, h{+}1) \quad h {<} n \quad \tau_{\infty} {=} sc(\sigma_{\infty},i)}
\\
\\
  \infer[(t{\cdot}r)_{\infty}]
  {cp_{\infty}(U, \sigma_{\infty}, h)}
  {cp_{\infty}(U, \tau_{\infty}, h{+}1) \quad h {<} n \quad \tau_{\infty} {=} mv(\sigma_{\infty},i,j)}
\end{array}
\]
The \emph{corecursive}\footnote{Corecursion is defined in Section
  \ref{sec:coind}. Note that these functions would be \emph{recursive}
  working with finite configurations.}  functions $zr,sc,mv:
Cgn_{\infty} \times \N^+ (\times \N^+) \to Cgn_{\infty}$ alter the
configurations, as prescribed by the instructions Zero, Successor and
Transfer; the definition of $zr$ is \eg\ as follows\footnote{We use
  here the notation $r : \sigma_{\infty}$ to represent the
  configuration $(0 {\mapsto} r,\ \iota{\mapsto}
  r_{\iota})^\rangeiMtwoinf$.}:
\[
\begin{array}{lcl}
zr(\sigma_{\infty},i) & \triangleq & \textrm{match $\sigma_{\infty}$ with $r : \tau_{\infty}$ $\Rightarrow$ match $i{-}1$ with }
                                      \textrm{$0$ $\Rightarrow$ $0 : \tau_{\infty}$ $|$ 
                                              $n{+}1$ $\Rightarrow$ $r : zr(\tau_{\infty},i{-}1)$}\\
\end{array}
\]
\end{adef}

The intended meaning of the judgment $cp_+(U,\sigma_{\infty},h,
\tau_{\infty})$ is that the computation under the program $U$ with the
configuration $\sigma_{\infty}$ and by starting from the $h$th
istruction of $U$, \emph{stops}, transforming $\sigma_{\infty}$ into
$\tau_{\infty}$.

On the other hand, the intended meaning of
$cp_{\infty}(U,\sigma_{\infty},h)$ is the same as $cp_{j\infty}$, even
if the configurations may be updated, in the case: the computation
under the program $U$ with the configuration $\sigma_{\infty}$ and by
starting from the $h$th istruction, \emph{loops}. That is, there
exists an instruction $I_q$ which can be reached from $I_h$ and such
that, afterwards, the control flow comes again at $I_q$ after a
non-zero, finite number of steps.
Nevertheless, the use of $cp_{\infty}$ is subtler than that of
$cp_{j\infty}$: the coinductive hypothesis (``and so on forever'') may
be actually applied, to grasp the divergence, \emph{provided} the
configuration at hand satisfies an \emph{invariant} (whose nature will
be clarified below).  Coherently with such an intuition, a
\emph{final} configuration (corresponding to the fourth parameter of
the inductive predicate $cp_+$) \emph{cannot} exist for $cp_{\infty}$,
simply because the configurations may be updated ``ad infinitum'' in
the course of a diverging computation!

Termination and divergence are now fully significant, and managed
parameterically as follows.

\begin{adef}\label{programeval}\;(Converging and diverging evaluation)
  Let $U$ and $\sigma_{\infty}$ be a program and a configuration, and
  let ${\mathcal T}(\sigma_{\infty},U)$, ${\mathcal
    I}(\sigma_{\infty},U)$ be decidable constraints about the content
  of the registers in $\sigma_{\infty}$, depending on $U$.  Then, the
  computation under $U$ with $\sigma_{\infty}$ \emph{converges} and
  \emph{diverges} when, respectively:
\[
\begin{array}[t]{lll}
  stop(U,\sigma_{\infty}) & \triangleq  & \exists \tau_{\infty},\ \exists \mathcal{T}(\sigma_{\infty},U).\ 
                                          \mathcal{T}(\sigma_{\infty},U) \Rightarrow cp_+(U, \sigma_{\infty}, 1, \tau_{\infty}) \\%[1mm]
  loop(U,\sigma_{\infty}) & \triangleq  & \exists \mathcal{I}(\sigma_{\infty},U).\ 
                                          \mathcal{I}(\sigma_{\infty},U) \Rightarrow cp_{\infty}(U, \sigma_{\infty}, 1)
\end{array}
\]
\end{adef}

As foreseen by the above comments about $cp_+$ and $cp_{\infty}$, the
management of convergence and divergence are fairly different between
each other, when the configurations can be updated by computations.

Converging computations under $U$ with initial $\sigma_{\infty}$ are
actually accommodated in the intuitive way: the halting is described
by the program counter, which is eventually set to $0$; moreover, the
incremental modification of $\sigma_{\infty}$ is reported in the final
$\tau_{\infty}$. The premise $\mathcal{T}(\sigma_{\infty},U)$ plays
the role of a \emph{termination} condition, which, if needed, provides
with the extra potential of carrying out proofs by induction.
In fact, computations may converge essentially in two ways: with or
without the presence of \emph{finite cycles}.
In the latter case, the constraint just ``guides'' the control flow to
the end of the program; in the presence of cycles, it is exploited to
pick out a parameter on which to reason by induction.
Therefore, in our \emph{logical} setting, program-driven termination
constraints make feasible formal proofs about the convergence and the
output of individual programs \wrt\ parameter configurations. In other
words, such conditions allow to make formal the informal proofs by
evidence that one may figure out by inspecting the programs.

Conversely, the modification of the starting configuration
$\sigma_{\infty}$ within diverging computations under $U$ does not
produce a final configuration, because $\sigma_{\infty}$ is updated ad
infinitum.
Though, the modification of $\sigma_{\infty}$ \emph{can be observed}
in the course of the computation, and such configuration may be
checked against an \emph{invariance} condition, that constrains its
content.  Therefore, the invariance condition
$\mathcal{I}(\sigma_{\infty},U)$ itself, whose shape depends again on
$U$, becomes the ``guard'' to ensure the non-termination\footnote{We
  remark that the whole scenario is coherent \wrt\ the concept of
  \emph{computable function}, that we will address in Section
  \ref{sec:minus}: there is an output, which is extracted from the
  final $\tau_{\infty}$, if and only if a computation stops.}.

Concerning the termination and invariance constraints, we restrict to
universally quantified formulas on natural numbers, built via the
logical operators and the arithmetic operations and predicates.

For the sake of illustrating the technical details, let us consider
the \emph{parametric} (\ie\ variable-content) configuration
$\mu_{\infty} {\triangleq} (1 {\mapsto} m,\ 2 {\mapsto} n,\ 3
{\mapsto} p, \ldots)$ and the program $V {\triangleq} \langle 1
{\mapsto} S(1),\ 2 {\mapsto} J(2,3,1) \rangle$.
We can then show that the computation under $V$ with $\mu_{\infty}$
diverges, by choosing the invariant $n {=} p$ (while it converges with
the termination constraint $n {\neq} p$).
To prove $\forall m,n,p.\ (n {=} p) \Rightarrow cp_{\infty}(V,
\mu_{\infty}, 1)$ by structural coinduction on the derivation within
\coq's top-down proof environment, we assume in the proof context the
coinductive hypothesis, the variables and the invariant; then we
execute the two instructions of $U$ so that the control flow loops
back to the first instruction; finally we apply the coinductive
hypothesis\footnote{The application of the coinductive hypothesis is
  \emph{guarded} by the two constructors $(s{\cdot}r)_{\infty}$ and
  $(jt{\cdot}r)_{\infty}$ (see also Section \ref{sec:coind}).}, which
demands to prove that the new configuration satisfies the invariant
constraint as well:
\[
  \infer[(1)]
  {\forall m,n,p {\in} \N.\ (n {=} p) \Rightarrow cp_{\infty}(U, (1 {\mapsto} m,\ 2 {\mapsto} n,\ 3 {\mapsto} p, \ldots), 1)}
    {\infer[(introduction)]
      {\forall m,n,p {\in} \N.\ (n {=} p) \Rightarrow cp_{\infty}(U, (1 {\mapsto} m,\ 2 {\mapsto} n,\ 3 {\mapsto} p, \ldots), 1)}
      {\infer[(s{\cdot}r)_{\infty}]
        {cp_{\infty}(U, (1 {\mapsto} m,\ 2 {\mapsto} n,\ 3 {\mapsto} p, \ldots), 1)}
        {\infer[(jt{\cdot}r)_{\infty}]
          {cp_{\infty}(U, (1 {\mapsto} m {+} 1,\ 2 {\mapsto} n,\ 3 {\mapsto} p, \ldots), 2)}
          {\infer[]
            {[cp_{\infty}(U, (1 {\mapsto} m {+} 1,\ 2 {\mapsto} n,\ 3 {\mapsto} p, \ldots), 1)]_{(1)}}
            {\infer*
              {n {=} p}
              {[n {=} p]}}
          }}}}
\]

\paragraph{\textbf{Adequacy (II).}}

We complete now the discussion about the faithfulness of our encoding
\wrt\ Cutland's URM \cite{cutland}, undertaken in Section
\ref{sec:urm}: the issues we have to address formally are the
relationship between finite and infinite configurations, and the
semantics given in the current and the previous section.

As far as the configurations are concerned, we first define the
\emph{inclusion} and \emph{restriction} concepts.

\begin{adef}\label{ext}\;(Configuration
  inclusion/restriction) Let $U$ be a program, $\sigma {=} (\iota
  {\mapsto} s_\iota)^\rangeiM$ a finite configuration and
  $\tau_{\infty} {=} (\iota {\mapsto} t_\iota)^\rangeiMinf$ an
  infinite one. Then, \emph{inclusion} and \emph{restriction} are
  defined as follows:
\[
\begin{array}{lcl}
  \sigma \subset \tau_{\infty} & \triangleq & (\forall \iota {\in} [1..m].\ t_\iota {=} s_\iota)\land
                                              (\forall \iota {>} m.\ t_\iota {=} 0)
\\
  \tau_{\infty | U}           & \triangleq & (\iota {\mapsto} t_\iota)^\rangeiRho
\end{array}
\]
\end{adef}

Concerning the semantics, let us assume (without displaying the rules)
to have introduced a second definition fot both the predicates $cp_+$
and $cp_{\infty}$, to cope with \emph{finite} configurations and for
which we use an overloaded notation.
The new rules differ from Definition \ref{eval} only for the fact that
the involved finite configurations require the extra compatibility
constraint with programs, analogously to Definition \ref{abs}.

Now we can state the \emph{equivalence} between finite and infinite
configurations encodings $Cgn$ and $Cgn_{\infty}$.

\begin{thm}\;(Configurations equivalence) Let $U {=} \langle \iota
  {\mapsto} I_\iota\rangle^\rangeiN$ be a program, $\sigma$ and $\tau$
  finite configurations, $\sigma_{\infty}$ and $\tau_{\infty}$
  infinite configurations, and let $h {\in} [1..n]$.  Then the
  following properties hold:
\begin{enumerate}
\item $cp_+(U,\sigma,h,\tau) \land \sigma {\models} U \land \sigma
  {\subset} \sigma_{\infty} \land \tau {\subset} \tau_{\infty}
  \Rightarrow cp_+(U,\sigma_{\infty},h,\tau_{\infty})$

\item $cp_{\infty}(U,\sigma,h) \land \sigma {\models} U \land \sigma
  {\subset} \sigma_{\infty} \Rightarrow
  cp_{\infty}(U,\sigma_{\infty},h)$

\item $cp_+(U,\sigma_{\infty},h,\tau_{\infty}) \Rightarrow
       cp_+(U,\sigma_{\infty | U},h,\tau_{\infty | U})$

\item $cp_{\infty}(U,\sigma_{\infty},h) \Rightarrow
       cp_{\infty}(U,\sigma_{\infty | U},h)$
\end{enumerate}
\proof (1, 3) By induction on the evaluation hypothesis. (2, 4) By
coinduction on the derivation.  
\end{thm}

Even if the above Theorem establishes that working either with finite,
list-like configurations or with infinite, stream-like ones, is
equivalent, we have preferred up to date to handle \emph{infinite}
configurations.  Our choice is motivated by two reasons: stream
configurations do not require the overhead of managing side-conditions
to model the compatibility with programs, and it has not been yet
necessary to perform proofs by induction on the structure of
configurations themselves.

In the end, the reader can see that our machinery provides the user
with a \emph{logic} for the URM, \ie\ a formal system whose potential
may be exploited to prove properties about the semantics of URM
programs and \emph{the encoding} itself, a direction we will comment
on further in the final section.

To consider the adequacy issue, we conjecture that our formalization
internalizes faithfully the very initial theory developed by Cutland
on paper, \ie\ the part concerning the synthesis and the execution of
individual programs.
By addressing the task formally, the \emph{soundness} of our encoding
is apparent (as our programs coincide with Cutland's ones, and we have
coupled to programs a formal logical system); moreover, we state a
limited form of \emph{completeness}, in the following sense.

\begin{cjt}\;(Adequacy) Let $P$ be an URM program and $U {=} \langle i
  {\mapsto} I_i\rangle^\rangeiN$ its faithful encoding.  Then:
\begin{enumerate}
\item If $P(a_1, a_2, \ldots, a_m) \!\!\downarrow\!  b$, then there exist $\tau  {=} \langle 1 {\mapsto} b, \iota{\mapsto} \tau_{\iota}\rangle^\rangeiMtwo$
      and $\mathcal{T}((\iota {\mapsto}a_\iota)^\rangeiM,U)$ such that
      $\mathcal{T}((\iota {\mapsto}a_\iota)^\rangeiM,U) \Rightarrow cp_+(U,(\iota {\mapsto}a_\iota)^\rangeiM,1,\tau)$

\item If $P(a_1, a_2, \ldots, a_m) \!\!\uparrow$, then there exists $\mathcal{I}((\iota {\mapsto}a_\iota)^\rangeiM,U)$ such that
      $\mathcal{I}((\iota {\mapsto}a_\iota)^\rangeiM,U) \Rightarrow cp_{\infty}(U,(\iota {\mapsto}a_\iota)^\rangeiM,1)$
\end{enumerate}
\proof (1) By inspection on the hypothetical evaluation (to devise the
termination constraint, which depends on the initial configuration
$(\iota_i {\mapsto}a_i)^\rangeiM$), then by induction (see also
Section \ref{sec:minus}).
(2) By inspection on the hypothetical evaluation (to devise the
invariant), then by structural coinduction.  
\end{cjt}

To conclude, we remark that, after the introduction of the very basic
computability theory, Cutland develops ``higher-order'' methods, to
devise new computable functions \emph{without} having to write
programs. It is immediate that addressing this kind of adequacy, at
the moment, is out of the scope of our approach.

%%%%%%%%%%%%%%%%%%%%%%%%%%%%%%%%%%%%%%%%%%%%%%%%%%%%%%%%%%%%%%%%

\section{An example: partial minus}
\label{sec:minus}

The next step of our work is to address slightly more involved
concepts: in this section we exploit the formalization developed so
far, by tuning it to deal with the \emph{functions} computed by the
URM.

The formal notion of \emph{(partial) computable function} arises
naturally in Cutland's presentation \cite{cutland} after the
preliminary definitions reported in Section \ref{sec:urm}. Namely, a
program $P$ computes a function $f{:}\;\N^m \rightharpoonup \N$ when,
for every $a_1, a_2, \ldots , a_m, b {\in} \N^{m+1}$, the computation
$P(a_1, a_2, \ldots , a_m)$ stops and $b$ is stored in the register
$R_1$ in the final configuration (this is written $P(a_1, a_2, \ldots\
, a_m)\!\downarrow\! b$) if and only if:
\[
(a_1, a_2, \ldots\ , a_m) {\in} dom(f) \ \textrm{and} \ f(a_1,
a_2, \ldots\ , a_m){=}b
\]

A relevant application supported by our machinery is to address the
\emph{certification} of URM programs: that is, proving that a program
meets the specification it is designed for.  
The example we will be working out in this section is the
\emph{partial} subtraction function $sub: \N {\times} \N
\rightharpoonup \N$:
\[
\begin{array}{lcl}
sub(m, n)  & \triangleq  & \left\{ \begin{array}{ll} m{-}n    & \quad \textrm{if $m {\geq} n$}\\
                                                     \uparrow & \quad \textrm{if $m {<} n$}
                                   \end{array}\right.
\end{array}
\]

An algorithm to make the URM compute this function is the following:
if $m$ and $n$ are loaded, respectively, in $R_1$ and $R_2$, then try
to let $n$ reach $m$ by performing \emph{Successor} operations on
$R_2$; correspondingly increment $R_3$, whose content is initially set
to $0$, to record the number of steps performed on $R_2$.  This
algorithm devises a \emph{loop} in the computation, which comes to an
end if and only if $m {\geq} n$. In any case, at any completion of the
loop, the snapshot of the registers content is the following:
\[
\begin{array}{ccccc}
R_1 & R_2     & R_3 & R_4 & \ldots
\\
m   & n {+} k & k   & 0   & \ldots
\end{array}
\]

The algortithm can be implemented, for example, by the following URM
program:
\[
U \triangleq \langle\ 1 {\mapsto} J(1,2,5),
                    \ 2 {\mapsto} S(2),
                    \ 3 {\mapsto} S(3),
                    \ 4 {\mapsto} J(1,1,1),
                    \ 5 {\mapsto} T(3,1)\ \rangle
\]

The program, as required, is designed to increment in parallel $r_2$
and $r_3$ and to stop just, and only if, when $r_2 {=} r_1$.
It is then immediate to see that the computations under $U$ may
converge or diverge depending on the initial configuration: therefore,
the implementation of the partial subtraction function has to be
certified in two steps, by using the predicates $cp_{\infty}$ and
$cp_+$ defined in the previous section.

On the one hand, we prove via $cp_{\infty}$ that the computation under
$U$ diverges with the configurations $(1 {\mapsto} m,\ 2 {\mapsto} n,
\ldots)$, such that $m {<} n$ (which is the ``invariant'').
To complete the analysis, we establish via $cp_+$ that the computation
under $U$ converges to $m{-}n$ with the configurations $(1 {\mapsto}
m,\ 2 {\mapsto} n,\ 3 {\mapsto} 0, \ldots)$, such that $m {\geq} n$
(this, in turn, plays the role of the ``termination'' constraint).

\begin{thm}\label{minus}\;(Partial minus)
  Let $\sigma {=} (1 {\mapsto} \sigma_1,\ 2 {\mapsto} \sigma_2,\ 3
  {\mapsto} \sigma_3, \ldots)$ be a parameter configuration.  Then,
  the implementation of the partial minus function is certified by the
  following properties:
\begin{enumerate}
\item (Divergence) $\sigma_1 {<} \sigma_2 \Rightarrow cp_{\infty}(U,\sigma,1)$

\item (Convergence) $\sigma_1 {\geq} \sigma_2 \Rightarrow cp_+(U, \sigma, 1, ( 1 {\mapsto} \sigma_1 {-} \sigma_2 {+} \sigma_3,\
                                                                               2 {\mapsto} \sigma_1,\
                                                                               3 {\mapsto} \sigma_1 {-} \sigma_2 {+} \sigma_3, \ldots) )$
\end{enumerate}
\proof (1.) By structural coinduction on the derivation. Assume the
coinductive hypothesis, then evaluate the first four instructions so
that the control flow loops back to the first instruction, finally
apply the coinductive hypothesis and prove that the updated
configuration satisfies the invariant constraint\footnote{See Section
  \ref{sec:coind} about the conventions for displaying \cccoind\
  top-down proofs in natural deduction style.}:
\[
  \infer[(1)]
  {\forall \sigma {=} (1 {\mapsto} \sigma_1,\ 2 {\mapsto} \sigma_2,\ 3 {\mapsto} \sigma_3, \ldots).\ 
    \sigma_1 {<} \sigma_2 \Rightarrow cp_{\infty}(U, \sigma, 1)}
  {\infer[(introduction)]
    {\forall \sigma {=} (1 {\mapsto} \sigma_1,\ 2 {\mapsto} \sigma_2,\ 3 {\mapsto} \sigma_3, \ldots).\ 
      \sigma_1 {<} \sigma_2 \Rightarrow cp_{\infty}(U, \sigma, 1)}
    {\infer[(jf{\cdot}r)_{\infty}]
      {cp_{\infty}(U, (1 {\mapsto} \sigma_1,\ 2 {\mapsto} \sigma_2,\ 3 {\mapsto} \sigma_3, \ldots), 1)}
      {\infer[(s{\cdot}r)_{\infty}]
        {cp_{\infty}(U, (1 {\mapsto} \sigma_1,\ 2 {\mapsto} \sigma_2,\ 3 {\mapsto} \sigma_3, \ldots), 2)}
        {\infer[(s{\cdot}r)_{\infty}]
          {cp_{\infty}(U, (1 {\mapsto} \sigma_1,\ 2 {\mapsto} \sigma_2 {+} 1,\ 3 {\mapsto} \sigma_3, \ldots), 3)}
          {\infer[(jt{\cdot}r)_{\infty}]
            {cp_{\infty}(U, (1 {\mapsto} \sigma_1,\ 2 {\mapsto} \sigma_2 {+} 1,\ 3 {\mapsto} \sigma_3 {+} 1, \ldots), 4)}
            {\infer[]
              {[cp_{\infty}(U, (1 {\mapsto} \sigma_1,\ 2 {\mapsto} \sigma_2 {+} 1,\ 3 {\mapsto} \sigma_3 {+} 1, \ldots), 1)]_{(1)}}
              {\infer*
                {\sigma_1 {<} \sigma_2 {+} 1}
                {[\sigma_1 {<} \sigma_2]}}
         }}}}}}
\]

(2.) By induction on $p {=} \sigma_1 {-} \sigma_2$. If $p {=} 0$, the
evaluation of the program $U$ reduces to obeying just the first
instruction (the Jump condition is true) and the last one, hence the
thesis is immediate.
If $p{=}q{+}1$, the evaluation of the first four instructions causes
the control flow to loop back to the first instruction, with the
configuration $(1 {\mapsto} \sigma_1,\ 2 {\mapsto} \sigma_2 {+} 1,\ 3
{\mapsto} \sigma_3 {+} 1, \ldots)$; the thesis follows from the
inductive hypothesis.

Finally, choosing $\sigma_3 {=} 0$ implies the convergence of the
computation under $U$ with $\sigma$ to $\sigma_1 {-} \sigma_2$.  
\end{thm}

\paragraph{\textbf{Inductive versus coinductive evaluations.}}

Regarding \emph{partial} functions, it is apparent that the two
predicates $cp_+$ and $cp_{\infty}$ act as complementary, being the
first one responsible for the treatment of the elements in the domain
of the function involved and the second one for all the extra
computations.

About this separation between inductive and \emph{purely} coinductive
evaluations, we wish to remark that it has not been possible to deal
with the semantics of URM programs by using a unique,
\emph{potentially} coinductive judgment. Actually, by restricting \eg\
on abstract programs, if such a predicate was defined through the
rules $(f{\cdot}l)_+$, $(t{\cdot}l)_+$, $(f{\cdot}r)_{\infty}$ and
$(t{\cdot}r)_{\infty}$ of Definition \ref{abs}, would be too weak.
Far from being an obstacle for our goals, this fact has caused just to
double a part of the encoding, to define both $cp_+$ and
$cp_{\infty}$; in any case, such a solution provides with an extra
proof principle, \ie\ the possibility of carrying out proofs by
structural induction on the derivation of converging computations.

Nevertheless, these considerations about the relationship between
inductive, potential and pure coinductive evaluation point out the
need of further research efforts, along the lines pursued by the much
more advanced work by Leroy and Grall \cite{leroy-grall09} (see the
next section for the discussion of related work).

%%%%%%%%%%%%%%%%%%%%%%%%%%%%%%%%%%%%%%%%%%%%%%%%%%%%%%%%%%%%%%%%

\section{Further and related work}
\label{sec:further}

In this document we have given an account of an experiment in
\cccoind, about modeling and reasoning on the execution of converging
and diverging low-level, assembly-like programs, carried out by the
Unlimited Register Machine (URM) \cite{cutland}.
The particular perspective which has inspired our research is the
formalization of a workbench to certify the implementation of the
functions computed by the URM; as a proof of concept, we have
addressed the partial minus function on natural numbers.
The encoding technique needed to accomplish our goal is quite plain,
apart from the use of the coinduction: in fact, we have taken most
advantage of the (co)inductive specification and proof principles
provided by the \cccoind\ intuitionistic type theory and mechanized in
the \coq\ proof assistant \cite{gimenez94,coq}.

In this final section we sketch some hints to exploit the potential of
our formalization, along two main directions: computability and traces
of execution.

\paragraph{\textbf{Computability.}}

In our work we have mastered the very basic computability theory of
the URM: essentially, we are able to prove that \emph{specific} URM
programs implement the functions they are designed for.  So we have
coupled a \emph{logic}, whose mechanization is supported by \coq, to
the bare URM. Nevertheless, exploiting the machinery requires a
non-trivial analysis and practice by the user, who has to pick out
ad-hoc properties (\emph{termination} and \emph{invariant} conditions)
to achieve the certification of URM code.

At this point, to pursue at a deeper extent the formalization of the
computability theory, one has to change a bit perspective, gaining a
more abstract level.
This opens actually two new directions, which form the core of the
computability: lifting from programs to functions (which they
implement) and describing ``higher-order'' methods, to combine such
functions for obtaining new, more sophisticated computable functions.
Therefore, one should add at least a new meta-level, where partial
functions are first-class citizens.
A possible approach towards this goal is to investigate more abstract
properties of URM programs, such as \emph{equivalence}. This effort,
in turn, would open further research lines, and tends again, as
invariance does, to the objective of capturing not only the outcome of
the execution of programs, but also the observable effects.

As far as we know, there is no related work about formalizing the
historical models used to develop the computability theory (and the
URM, in particular). We see this as a serious gap from the point of
view of certified mathematics, a framework where the research is
nowadays intense; hence the present document is also an effort to
contribute closing this gap.

\paragraph{\textbf{Traces of execution.}}

Leroy and Grall \cite{leroy-grall09} adopt coinduction within
\cccoind\ to capture both finite and infinite evaluations of a
\emph{call-by-value $\lambda$-calculus}.
The motivation of that work is the attempt to describe big-step
semantics by coinduction, because big-step semantics is more
convenient than small-step to prove the correctness of program
transformations, such as \emph{compilation}.
Nevertheless, big-step semantics is traditionally defined by
induction, thus allowing to describe only terminating evaluation.

Grall and Leroy prove that (only) a big-step semantics that separates
terminating evaluation (described by an inductive predicate) from
diverging evaluation (described by a purely coinductive predicate)
corresponds exactly to finite and non-finite small-step reductions.
Afterwards, the authors extend both the semantics to produce not only
the outcome of an evaluation (convergence and output, or divergence)
but also an \emph{execution trace}, in the form of a potentially
infinite sequence of terms representing the intermediate reducts of
the source program. This extension is fundamental to establish
semantic preservation properties for program transformation (such as
compilation) and is very important to investigate observational
equivalence for imperative languages.

Therefore, it would be stimulating to experiment with traces of
execution for the URM (for example in the form of potential infinite
sequences of configurations) to address \eg\ equivalence of programs.

\paragraph{\textbf{Other work related to divergence or low-level
    languages.}}

There are several contributions in the literature exploiting the
potential of coinductive definitions and proofs within \cccoind\ to
master the fundamental concept of non-terminating computation.
Some of these approaches concern transition systems
\cite{coupet-jacubiec99,bertot:book}, linear temporal logic
\cite{coupet03,bertot:book} and process algebras
\cite{gimenez95,pi01}.

Finally, from a complementary point of view, we observe that in recent
years the metatheory of low-level machines has been studied by several
authors in more realistic settings
\cite{crary03,tan-appel04,chlipala07}.

\bibliographystyle{eptcs}
\bibliography{infinity-11}

\begin{thebibliography}{10}
\providecommand{\bibitemdeclare}[2]{}
\providecommand{\urlprefix}{Available at }
\providecommand{\url}[1]{\texttt{#1}}
\providecommand{\href}[2]{\texttt{#2}}
\providecommand{\urlalt}[2]{\href{#1}{#2}}
\providecommand{\doi}[1]{doi:\urlalt{http://dx.doi.org/#1}{#1}}
\providecommand{\bibinfo}[2]{#2}

\bibitemdeclare{proceedings}{DBLP:conf/types/1993}
\bibitem{DBLP:conf/types/1993}
\bibinfo{editor}{H.~Barendregt} \& \bibinfo{editor}{T.~Nipkow}, editors
  (\bibinfo{year}{1994}): \emph{\bibinfo{title}{Types for Proofs and Programs,
  International Workshop TYPES'93, Nijmegen, The Netherlands, May 24-28, 1993,
  Selected Papers}}. {\sl \bibinfo{series}{Lecture Notes in Computer Science}}
  \bibinfo{volume}{806}, \bibinfo{publisher}{Springer}.

\bibitemdeclare{proceedings}{DBLP:conf/types/1995}
\bibitem{DBLP:conf/types/1995}
\bibinfo{editor}{S.~Berardi} \& \bibinfo{editor}{M.~Coppo}, editors
  (\bibinfo{year}{1996}): \emph{\bibinfo{title}{Types for Proofs and Programs,
  International Workshop TYPES'95, Torino, Italy, June 5-8, 1995, Selected
  Papers}}. {\sl \bibinfo{series}{Lecture Notes in Computer Science}}
  \bibinfo{volume}{1158}, \bibinfo{publisher}{Springer}.

\bibitemdeclare{book}{bertot:book}
\bibitem{bertot:book}
\bibinfo{author}{Y.~Bertot} \& \bibinfo{author}{P.~Cast{\'e}ran}
  (\bibinfo{year}{2004}): \emph{\bibinfo{title}{Interactive Theorem Proving and
  Program Development, Coq'Art:the Calculus of Inductive Constructions}}.
\newblock \bibinfo{publisher}{Springer-Verlag}.

\bibitemdeclare{proceedings}{DBLP:conf/tphol/1999}
\bibitem{DBLP:conf/tphol/1999}
\bibinfo{editor}{Y.~Bertot}, \bibinfo{editor}{G.~Dowek},
  \bibinfo{editor}{A.~Hirschowitz}, \bibinfo{editor}{C.~Paulin} \&
  \bibinfo{editor}{L.~Th{\'e}ry}, editors (\bibinfo{year}{1999}):
  \emph{\bibinfo{title}{Theorem Proving in Higher Order Logics, 12th
  International Conference, TPHOLs'99, Nice, France, September, 1999,
  Proceedings}}. {\sl \bibinfo{series}{Lecture Notes in Computer Science}}
  \bibinfo{volume}{1690}, \bibinfo{publisher}{Springer}.

\bibitemdeclare{inproceedings}{bertot:reccorec}
\bibitem{bertot:reccorec}
\bibinfo{author}{Y.~Bertot} \& \bibinfo{author}{E.~Komendantskaya}
  (\bibinfo{year}{2009}): \emph{\bibinfo{title}{Using Structural Recursion for
  Corecursion}}.
\newblock In \bibinfo{editor}{S.~Berardi}, \bibinfo{editor}{F.~Damiani} \&
  \bibinfo{editor}{U~de'Liguoro}, editors: {\sl \bibinfo{booktitle}{Types for
  proofs and programs 2008}}, {\sl \bibinfo{series}{Lecture Notes in Computer
  Science}} \bibinfo{volume}{5497}, \bibinfo{publisher}{Springer}, pp.
  \bibinfo{pages}{220--236}.
\newblock \urlprefix\url{http://hal.inria.fr/inria-00322331}.

\bibitemdeclare{inproceedings}{chlipala07}
\bibitem{chlipala07}
\bibinfo{author}{A.~Chlipala} (\bibinfo{year}{2007}): \emph{\bibinfo{title}{A
  certified type-preserving compiler from lambda calculus to assembly
  language}}.
\newblock In \bibinfo{editor}{Ferrante} \& \bibinfo{editor}{McKinley}
  \cite{DBLP:conf/pldi/2007}, pp. \bibinfo{pages}{54--65}.
\newblock \urlprefix\url{http://doi.acm.org/10.1145/1250734.1250742}.

\bibitemdeclare{misc}{alberto:url}
\bibitem{alberto:url}
\bibinfo{author}{A.~Ciaffaglione} (\bibinfo{year}{2011}):
  \emph{\bibinfo{title}{The Web Appendix of this paper}}.
\newblock \urlprefix\url{http://www.dimi.uniud.it/ciaffagl}.

\bibitemdeclare{inproceedings}{coquand93}
\bibitem{coquand93}
\bibinfo{author}{T.~Coquand} (\bibinfo{year}{1993}):
  \emph{\bibinfo{title}{Infinite Objects in Type Theory}}.
\newblock In \bibinfo{editor}{Barendregt} \& \bibinfo{editor}{Nipkow}
  \cite{DBLP:conf/types/1993}, pp. \bibinfo{pages}{62--78}.
\newblock \urlprefix\url{http://dx.doi.org/10.1007/3-540-58085-9_72}.

\bibitemdeclare{article}{coupet03}
\bibitem{coupet03}
\bibinfo{author}{S.~Coupet-Grimal} (\bibinfo{year}{2003}):
  \emph{\bibinfo{title}{An Axiomatization of Linear Temporal Logic in the
  Calculus of Inductive Constructions}}.
\newblock {\sl \bibinfo{journal}{J. Log. Comput.}}
  \bibinfo{volume}{13}(\bibinfo{number}{6}), pp. \bibinfo{pages}{801--813}.
\newblock \urlprefix\url{http://dx.doi.org/10.1093/logcom/13.6.801}.

\bibitemdeclare{inproceedings}{coupet-jacubiec99}
\bibitem{coupet-jacubiec99}
\bibinfo{author}{S.~Coupet-Grimal} \& \bibinfo{author}{L.~Jakubiec}
  (\bibinfo{year}{1999}): \emph{\bibinfo{title}{Hardware Verification Using
  Co-induction in COQ}}.
\newblock In \bibinfo{editor}{\bibinfo{editor}{Bertot}} et~al.
  \cite{DBLP:conf/tphol/1999}, pp. \bibinfo{pages}{91--108}.
\newblock \urlprefix\url{http://dx.doi.org/10.1007/3-540-48256-3_7}.

\bibitemdeclare{inproceedings}{crary03}
\bibitem{crary03}
\bibinfo{author}{K.~Crary} (\bibinfo{year}{2003}):
  \emph{\bibinfo{title}{Toward a foundational typed assembly language}}.
\newblock In: {\sl \bibinfo{booktitle}{POPL}}, pp. \bibinfo{pages}{198--212}.
\newblock \urlprefix\url{http://doi.acm.org/10.1145/640128.604149}.

\bibitemdeclare{book}{cutland}
\bibitem{cutland}
\bibinfo{author}{N.~J. Cutland} (\bibinfo{year}{1980}):
  \emph{\bibinfo{title}{Computability: An Introduction to Recursive Function
  Theory}}.
\newblock \bibinfo{publisher}{Cambridge University Press}.

\bibitemdeclare{proceedings}{DBLP:conf/types/1994}
\bibitem{DBLP:conf/types/1994}
\bibinfo{editor}{P.~Dybjer}, \bibinfo{editor}{B.~Nordstr{\"o}m} \&
  \bibinfo{editor}{J.~M. Smith}, editors (\bibinfo{year}{1995}):
  \emph{\bibinfo{title}{Types for Proofs and Programs, International Workshop
  TYPES'94, B{\aa}stad, Sweden, June 6-10, 1994, Selected Papers}}. {\sl
  \bibinfo{series}{Lecture Notes in Computer Science}} \bibinfo{volume}{996},
  \bibinfo{publisher}{Springer}.

\bibitemdeclare{proceedings}{DBLP:conf/pldi/2007}
\bibitem{DBLP:conf/pldi/2007}
\bibinfo{editor}{J.~Ferrante} \& \bibinfo{editor}{K.~S. McKinley},
  editors (\bibinfo{year}{2007}): \emph{\bibinfo{title}{Proceedings of the ACM
  SIGPLAN 2007 Conference on Programming Language Design and Implementation,
  San Diego, California, USA, June 10-13, 2007}}. \bibinfo{publisher}{ACM}.

\bibitemdeclare{proceedings}{DBLP:conf/types/2002}
\bibitem{DBLP:conf/types/2002}
\bibinfo{editor}{H.~Geuvers} \& \bibinfo{editor}{F.~Wiedijk}, editors
  (\bibinfo{year}{2003}): \emph{\bibinfo{title}{Types for Proofs and Programs,
  Second International Workshop, TYPES 2002, Berg en Dal, The Netherlands,
  April 24-28, 2002, Selected Papers}}. {\sl \bibinfo{series}{Lecture Notes in
  Computer Science}} \bibinfo{volume}{2646}, \bibinfo{publisher}{Springer}.

\bibitemdeclare{inproceedings}{digianantonio-miculan02}
\bibitem{digianantonio-miculan02}
\bibinfo{author}{P.~Di Gianantonio} \& \bibinfo{author}{M.~Miculan}
  (\bibinfo{year}{2002}): \emph{\bibinfo{title}{A Unifying Approach to
  Recursive and Co-recursive Definitions}}.
\newblock In \bibinfo{editor}{Geuvers} \& \bibinfo{editor}{Wiedijk}
  \cite{DBLP:conf/types/2002}, pp. \bibinfo{pages}{148--161}.
\newblock \urlprefix\url{http://dx.doi.org/10.1007/3-540-39185-1_9}.

\bibitemdeclare{inproceedings}{gimenez94}
\bibitem{gimenez94}
\bibinfo{author}{E.~Gim{\'e}nez} (\bibinfo{year}{1994}):
  \emph{\bibinfo{title}{Codifying Guarded Definitions with Recursive Schemes}}.
\newblock In \bibinfo{editor}{\bibinfo{editor}{Dybjer}} et~al.
  \cite{DBLP:conf/types/1994}, pp. \bibinfo{pages}{39--59}.
\newblock \urlprefix\url{http://dx.doi.org/10.1007/3-540-60579-7_3}.

\bibitemdeclare{inproceedings}{gimenez95}
\bibitem{gimenez95}
\bibinfo{author}{E.~Gim{\'e}nez} (\bibinfo{year}{1995}):
  \emph{\bibinfo{title}{An Application of Co-inductive Types in Coq:
  Verification of the Alternating Bit Protocol}}.
\newblock In \bibinfo{editor}{Berardi} \& \bibinfo{editor}{Coppo}
  \cite{DBLP:conf/types/1995}, pp. \bibinfo{pages}{135--152}.
\newblock \urlprefix\url{http://dx.doi.org/10.1007/3-540-61780-9_67}.

\bibitemdeclare{inproceedings}{gimenez98}
\bibitem{gimenez98}
\bibinfo{author}{E.~Gim{\'e}nez} (\bibinfo{year}{1998}):
  \emph{\bibinfo{title}{Structural Recursive Definitions in Type Theory}}.
\newblock In \bibinfo{editor}{\bibinfo{editor}{Larsen}} et~al.
  \cite{DBLP:conf/icalp/1998}, pp. \bibinfo{pages}{397--408}.
\newblock \urlprefix\url{http://dx.doi.org/10.1007/BFb0055070}.

\bibitemdeclare{article}{pi01}
\bibitem{pi01}
\bibinfo{author}{F.~Honsell}, \bibinfo{author}{M.~Miculan} \&
  \bibinfo{author}{I.~Scagnetto} (\bibinfo{year}{2001}):
  \emph{\bibinfo{title}{pi-calculus in (Co)inductive-type theory}}.
\newblock {\sl \bibinfo{journal}{Theor. Comput. Sci.}}
  \bibinfo{volume}{253}(\bibinfo{number}{2}), pp. \bibinfo{pages}{239--285}.
\newblock \urlprefix\url{http://dx.doi.org/10.1016/S0304-3975(00)00095-5}.

\bibitemdeclare{proceedings}{DBLP:conf/icalp/1998}
\bibitem{DBLP:conf/icalp/1998}
\bibinfo{editor}{K.~Guldstrand Larsen}, \bibinfo{editor}{S.~Skyum} \&
  \bibinfo{editor}{G.~Winskel}, editors (\bibinfo{year}{1998}):
  \emph{\bibinfo{title}{Automata, Languages and Programming, 25th International
  Colloquium, ICALP'98, Aalborg, Denmark, July 13-17, 1998, Proceedings}}. {\sl
  \bibinfo{series}{Lecture Notes in Computer Science}} \bibinfo{volume}{1443},
  \bibinfo{publisher}{Springer}.

\bibitemdeclare{article}{leroy-grall09}
\bibitem{leroy-grall09}
\bibinfo{author}{X.~Leroy} \& \bibinfo{author}{Herv{\'e} Grall}
  (\bibinfo{year}{2009}): \emph{\bibinfo{title}{Coinductive big-step
  operational semantics}}.
\newblock {\sl \bibinfo{journal}{Inf. Comput.}}
  \bibinfo{volume}{207}(\bibinfo{number}{2}), pp. \bibinfo{pages}{284--304}.
\newblock \urlprefix\url{http://dx.doi.org/10.1016/j.ic.2007.12.004}.

\bibitemdeclare{article}{urm63}
\bibitem{urm63}
\bibinfo{author}{J.~C. Shepherdson} \& \bibinfo{author}{H.~E. Sturgis}
  (\bibinfo{year}{1963}): \emph{\bibinfo{title}{Computability of Recursive
  Functions}}.
\newblock {\sl \bibinfo{journal}{J. ACM}}
  \bibinfo{volume}{10}(\bibinfo{number}{2}), pp. \bibinfo{pages}{217--255}.
\newblock \urlprefix\url{http://doi.acm.org/10.1145/321160.321170}.

\bibitemdeclare{proceedings}{DBLP:conf/vmcai/2004}
\bibitem{DBLP:conf/vmcai/2004}
\bibinfo{editor}{B.~Steffen} \& \bibinfo{editor}{G.~Levi}, editors
  (\bibinfo{year}{2004}): \emph{\bibinfo{title}{Verification, Model Checking,
  and Abstract Interpretation, 5th International Conference, VMCAI 2004,
  Venice, January 11-13, 2004, Proceedings}}. {\sl \bibinfo{series}{Lecture
  Notes in Computer Science}} \bibinfo{volume}{2937},
  \bibinfo{publisher}{Springer}.

\bibitemdeclare{inproceedings}{tan-appel04}
\bibitem{tan-appel04}
\bibinfo{author}{G.~Tan}, \bibinfo{author}{A.~W. Appel},
  \bibinfo{author}{K.~N. Swadi} \& \bibinfo{author}{D.~Wu}
  (\bibinfo{year}{2004}): \emph{\bibinfo{title}{Construction of a Semantic
  Model for a Typed Assembly Language}}.
\newblock In \bibinfo{editor}{Steffen} \& \bibinfo{editor}{Levi}
  \cite{DBLP:conf/vmcai/2004}, pp. \bibinfo{pages}{30--43}.
\newblock \urlprefix\url{http://dx.doi.org/10.1007/978-3-540-24622-0_4}.

\bibitemdeclare{manual}{coq}
\bibitem{coq}
\bibinfo{author}{The Coq~Development Team} (\bibinfo{year}{2010}):
  \emph{\bibinfo{title}{The Coq Proof Assitant Reference Manual, version 8.3}}.
\newblock \bibinfo{organization}{INRIA}.

\end{thebibliography}

\end{document}